# Novel mid-infrared dispersive wave generation in gas-filled PCF by transient ionization-driven changes in dispersion

F. Köttig[1*], D. Novoa[1], F. Tani[1], M. Cassataro[1], J. C. Travers[1,2], and P. St.J. Russell[1]

[1]Max Planck Institute for the Science of Light, Staudtstrasse 2, 91058 Erlangen, Germany

[2]School of Engineering and Physical Sciences, Heriot-Watt University, Edinburgh EH14 4AS, U.K.

[*]e-mail: felix.koettig@mpl.mpg.de

**Abstract**

Gas-filled hollow-core photonic crystal fibre (PCF) is being used to generate ever wider supercontinuum spectra, in particular via dispersive wave (DW) emission in the deep and vacuum ultraviolet, with a multitude of applications. DWs are the result of the resonant transfer of energy from a self-compressed soliton, a process which relies crucially on phase matching. It was recently predicted that, in the strong-field regime, the additional transient anomalous dispersion introduced by gas ionization would allow phase-matched DW generation in the mid-infrared (MIR)—something that is forbidden in the absence of free electrons. Here we report for the first time the experimental observation of such MIR DWs, embedded in a 4.7-octave-wide supercontinuum that uniquely reaches simultaneously to the vacuum ultraviolet, with up to 1.7 W of total average power.

**Introduction**

Coherent nonlinear interactions between intense laser light and photo-induced plasmas are important in fields such as optical filamentation[1], laser wakefield acceleration[2], attoscience (through high harmonic generation[3]) and wideband terahertz pulse generation[4]. Recently there has been increased interest in the development of coherent ultrafast sources in the MIR, driven by applications in time-resolved molecular spectroscopy[5] and generation of small-footprint ultrafast x-ray lasers[6].

Many efficient MIR sources are based on solid-state materials, employing techniques such as difference frequency generation[7], optical parametric amplification[8], and DW emission from solitons in crystals with cascaded quadratic nonlinearities[9]. Due to the limited transparency range of the crystals, however, they cannot simultaneously provide tunable ultrashort pulses in the deep (DUV) and vacuum (VUV) ultraviolet—wavelength ranges where the majority of bio-polymers have electronic resonances (highly relevant for photoemission[10] and bio-spectroscopy[11]).

For these applications, gas-filled hollow-core PCF, guiding broad-band by anti-resonant reflection (ARR), has emerged as an excellent alternative, free of many of the limitations of established techniques[12]. ARR-PCFs offer long well-controlled collinear path-lengths and high damage thresholds, permitting operation even in the strong-field regime[13]. ARR-PCFs come in two main varieties—kagomé-type[14] and single-ring[15–17]—and offer weak anomalous waveguide dispersion that can be counter-balanced by the normal dispersion of the filling gas, enabling well-controlled bright soliton dynamics[18]. These characteristics make it possible to obtain single-cycle pulses with µJ energies via soliton self-compression[19,20]. Along with the compression process, ultrashort bunches of resonant radiation can be efficiently emitted as tunable DWs on the opposite side of the zero dispersion point (ZDP) in the normal dispersion regime[21–23].

Upon self-compression of the input pulse, the peak intensity reaches a level high enough to partially ionize the gas, allowing the hollow-core PCF system to be used for precise control of soliton-plasma interactions. The presence of free electrons remarkably enables resonant transfer of energy into MIR DWs in the anomalous dispersion region, i.e., on the same side of the ZDP as the pump[24]. This counterintuitive effect results from a plasma-induced transient change in the frequency dependence of the refractive index (normally assumed to be independent of optical intensity) that triggers phase-matched emission of MIR DWs.

Here we report the first experimental observation of this effect, resulting in the simultaneous emission of coherent MIR and DUV/VUV light. Intense nonlinear dynamics in the ARR-PCF creates a supercontinuum that is at least (limited by our detection system) 4.7 octaves (1.6 PHz) wide, spanning from 180 nm to 4.7 µm, with up to 1.7 W of average power.

**Results—Experimental results**

The underlying soliton-plasma interaction takes place in a noble-gas-filled kagomé-type ARR-PCF (36 µm core diameter, 7 cm long) which is pumped by pulses of duration 27 fs (full-width-half-maximum - FWHM) and central wavelength 1030 nm, with energies up to 16 µJ and a repetition rate of 151 kHz (Methods).

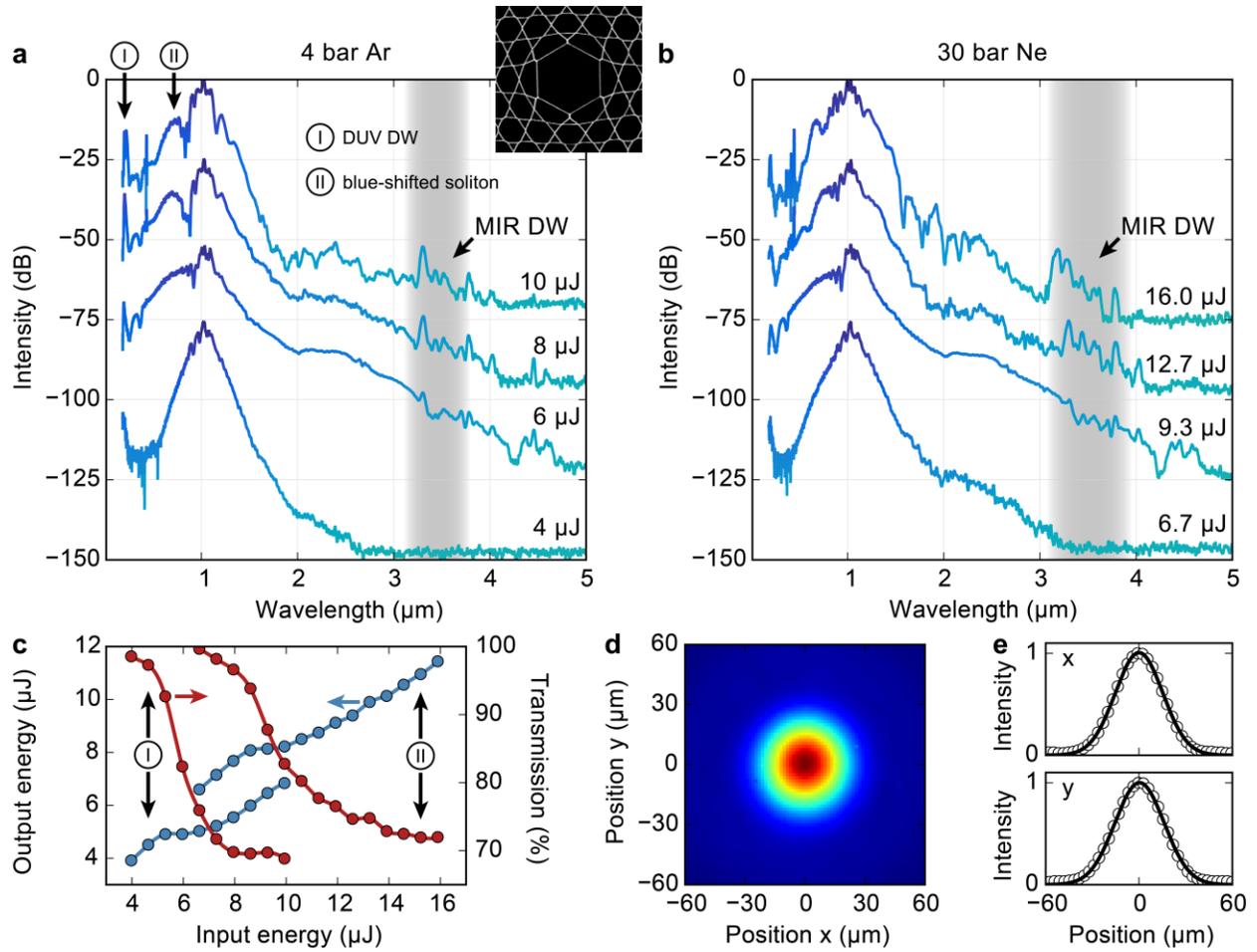

**Figure 1 | Experimental results. a,** Output spectra in the range 180 nm to 5 μm, when 7 cm of fibre was filled with 4 bar of Ar and pumped with 4 to 10 μJ pulses (the spectra are offset by 25 dB each for clarity). The spectra are not intensity-calibrated below 200 nm and above 900 nm. In the vicinity of 4.2 μm, absorption of carbon dioxide in the 2-m-long detection path in air reduced the signal, while water absorption was negligible for the spectral region of interest (the relative humidity of the laboratory air was ~30%). In numerical simulations (Methods) based on the experimental parameters in a and b, MIR DWs are emitted in the grey-shaded spectral regions. The inset shows the microstructure of the kagomé-PCF used in the experiment. **b,** Output spectra when the fibre was filled with 30 bar of Ne. **c,** Output energy and transmission of the fibre (including incoupling to the fibre). (I)—4 bar Ar. (II)—30 bar Ne. **d,** Optical near-field profile of the fibre mode for wavelengths longer than 3.1 μm, when the fibre was filled with 4 bar of Ar, and pumped with 6.6 μJ pulses. **e,** Lineouts in x- and y-direction through the centre of the mode. The dots are the measured points, and the lines are Gaussian fits.

When the fibre is filled with 4 bar of Ar, the ZDP is located at 517 nm and the 1030 nm pump pulses lie in the anomalous dispersion region where they are subject to soliton dynamics. For the available range of input energies (4 to 10 μJ) the soliton order lies between 3.3 and 5.2, well below the modulational instability regime[25], so that coherent soliton fission dominates the dynamics and high self-compression purity is achieved[26]. For input energies greater than 6 μJ, the higher-order input solitons strongly self-compress in the fibre, generating a broadband supercontinuum spanning from 180 nm to 4.7 μm (Fig. 1a). A DUV DW is emitted at a phase-matched wavelength of ~200 nm, with up to 35 nJ pulse energy (5.3 mW of average power, corresponding to 0.6% of the total output). According to numerical simulations (Methods), photoionization in the vicinity of the point of maximum temporal compression creates plasma densities exceeding $5 \cdot 10^{17}$ cm$^{-3}$, resulting in a transient change in the dispersion that gives rise to phase-matched generation of a second DW on the long-wavelength edge of the supercontinuum tail. The bandwidth of this MIR DW extends from 3.2 to 4.7 μm, with up to 1 nJ pulse energy (0.15 mW of average power, corresponding to 0.02% of the total output). Its spectral position is in good agreement with the numerical simulations (compare the grey-shaded area in Fig. 1a,b), immediately verifying the underlying phase matching mechanism (a plasma density of $5.5 \cdot 10^{17}$ cm$^{-3}$ yields phase matching to MIR DWs at 3.5 μm in both 4 bar of Ar and 30 bar of Ne). Note that a MIR spectrum extending from 3 to 4 μm is sufficient to support sub-3-cycle pulses with durations of 30 fs (Supplementary Section 4). The numerical simulations predict that the phase is spectrally flat[24]. Under these circumstances the peak power would exceed 20 kW for a transform-limited pulse. The MIR DW is also accompanied by blue-shifting solitons at wavelengths below 800 nm—a typical feature of pulse propagation in the presence of photoionization[13]. The losses associated with the generation of free electrons are apparent in the measured output energy (Fig. 1c), the transmission dropping strongly after the onset of soliton blue-shifting and MIR DW emission.

The MIR DW emission dynamics are governed by the subtle interplay between spectral broadening and compression of the input pulse, and plasma formation in the gas. In order to investigate the influence of the gas dispersion, the intensity at the self-compression point and the ionization dynamics, the fibre was filled with the lighter noble gas Ne, which has weaker dispersion and a higher ionization potential. For a filling pressure of 30 bar, the ZDP is 482 nm. For soliton orders between 3.2 and 5.0 (corresponding to input energies between 6.7 and 16 μJ), the dynamics were very similar to those seen in the Ar-filled fibre. Also, very distinct MIR DWs were emitted in this case (Fig. 1b), with pulse energies up to 1.3 nJ (0.2 mW of average power, corresponding to 0.016% of the total output). Despite the different gas type and pressure, and despite the higher input energies, powers and intensities, the MIR DWs in this experiment covered a spectral range similar to that for the Ar-filled fibre. This is due to the dominance of the plasma contribution to phase matching, and is analysed in more detail in the next section. Due to the weaker dispersion of Ne at short wavelengths, the DW phase matching wavelength shifts down to 140 nm in the VUV[23]. Owing to the complexity of measurements in the VUV, which must be performed under high vacuum, it was not possible with the existing set-up to make simultaneous measurements in the VUV. Instead we made use of a previous set of experiments (same system, same fibre, also filled with 30 bar of Ne, but with a length of 15 cm instead of 7 cm (Fig. 2)).

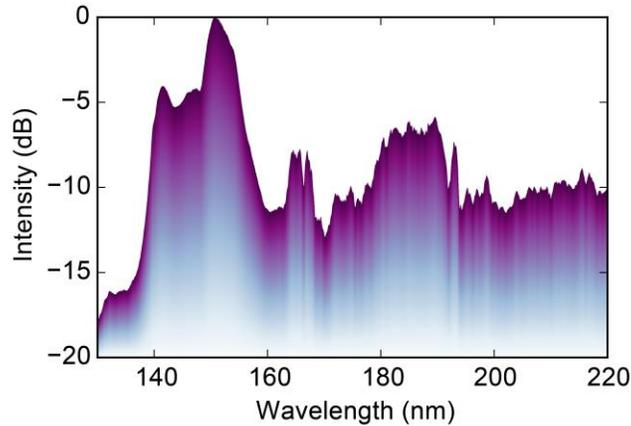

**Figure 2 | VUV measurement.** VUV spectrum (measured in a separate experiment; intensity-calibrated, pulse energy ~9 µJ) for 15 cm of an identical fibre filled with 30 bar of Ne.

For DW emission in the VUV a massive temporal shock is required (adiabatic pulse compression is insufficient), which for our experimental parameters occurs after only a few cm of propagation. It is therefore highly likely that VUV DWs are emitted also in the current experiment, a conclusion that is strongly supported by the observed dynamics (in particular the shock and the subsequent spectral recoil that generates the MIR DWs). Including these VUV measurements, the supercontinuum spectrum is 5.1 octaves (2.1 PHz) wide.

In both gases, the highest input pulse energies create significant plasma densities well before the self-compression point, strongly disrupting the self-compression process and resulting in a much more structured spectrum. The DUV and MIR DWs remain largely unaffected, however, which underlines their different origin, clearly distinguishing them from transient features in the output spectrum. Although the kagomé-PCF used in the experiment is not single-mode, careful launch alignment minimizes the excitation of higher-order modes. The MIR near-field profile of the fibre mode at the output face (Fig. 1d,e) has a near-perfect Gaussian shape, with no noticeable higher-order mode content. Note that potential light-glass interactions and resulting propagation losses in the MIR can be minimized by bringing the MIR DW emission point close to the output fibre end, either by optimizing the length of the fibre, or by simply tuning the input energy (the higher the input energy, the earlier the pulses compress upon propagation through the fibre).

Further experiments were performed in the same kagomé-PCF, with helium or krypton as filling gas, and also in a single-ring ARR-PCF (49 μm core diameter). In all cases, MIR DWs were emitted in the spectral range between 3 and 5 μm, which is in very good agreement with the numerical simulations and results from saturation of the plasma density to ~$10^{18}$ cm$^{-3}$.

## Results—Theory and simulation

Figure 3 shows the simulated pulse propagation (Methods), when the fibre is filled with 4 bar of Ar and pumped with 6 μJ pulses.

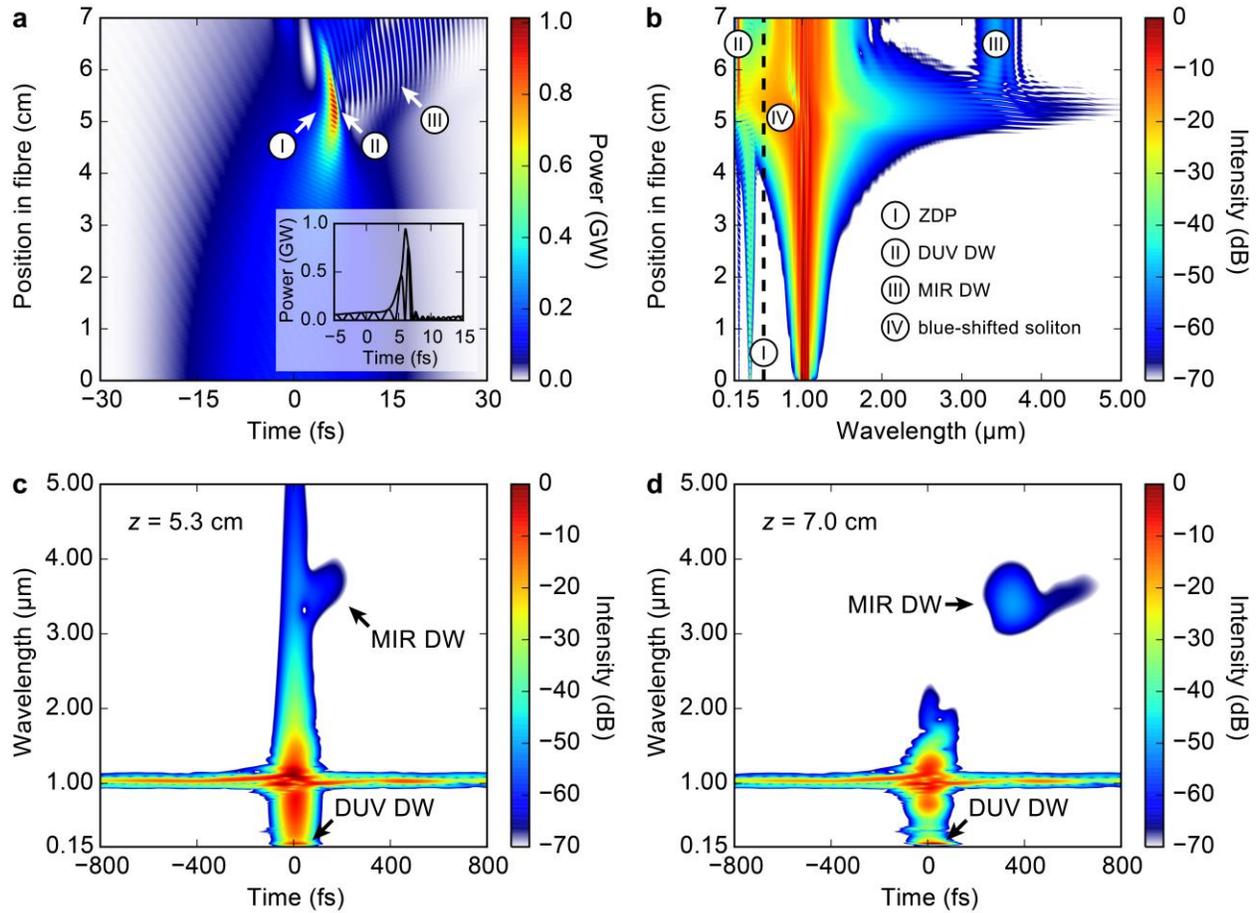

**Figure 3 | Simulated pulse propagation and time-frequency analysis.** Fibre filled with 4 bar of Ar, 6 μJ input pulses (measured using frequency-resolved optical gating (FROG)). **a,** Evolution of the temporal pulse shape with propagation distance. The inset shows the pulse at the maximum temporal compression point. (I)—maximum temporal compression point. (II)—shock front. (III)—DUV DW. **b,** Evolution of the normalized spectral intensity (per unit wavelength) with propagation distance. The dispersion is anomalous on the long-wavelength side of the ZDP. **c,** Spectrogram of the pulse at 5.3 cm (at the maximum temporal compression point), calculated using a 50 fs Gaussian gate pulse. The long-

duration low-intensity background in the vicinity of 1030 nm is the residual uncompressed low-intensity signal after nonlinear compression (Methods). **d,** Spectrogram of the pulse at 7 cm (fibre output).

At the maximum temporal compression point (Fig. 3a, ~5.3 cm) the soliton is self-compressed to a FWHM duration of 1.5 fs, with less than one optical cycle under the intensity envelope. The spectrogram of the pulse at this point (Fig. 3c) confirms soliton self-compression, yielding a near-transform-limited pulse. Upon further propagation, the optical shock effect at the trailing edge of the pulse strongly enhances the short-wavelength side of the spectrum, and energy is efficiently transferred to a phase-matched DW in the normal dispersion region at ~200 nm (Fig. 3b). At the maximum temporal compression point the peak power reaches the GW-level and the intensity rises above $2 \cdot 10^{14}$ W cm$^{-2}$, resulting in significant plasma densities in the non-perturbative strong-field regime (the Keldysh parameter[27] $\gamma \sim 0.6$ at this point). The spectral recoil that accompanies DUV DW emission enhances the long-wavelength side of the spectrum[28,29], and the generated plasma densities (of order $5 \cdot 10^{17}$ cm$^{-3}$) allow phase matching to a MIR DW in the wavelength range 3 to 4 μm, in the anomalous dispersion region. Due to its large group velocity mismatch with respect to the pump soliton, the MIR DW has detached from the main pulse by the time it reaches the fibre output (Fig. 3d), underlining its linear-wave character.

Phase matching between solitons and linear waves requires momentum conservation, i.e., $\Delta\beta(\omega) = \beta(\omega) - \beta_{sol}(\omega) = 0$, where $\omega$ is frequency, and $\beta$ and $\beta_{sol}$ are the propagation constants of linear waves and solitons. An analytical model that includes the contribution of plasma polarization to the nonlinear soliton propagation constant takes the form[24]:

$$\Delta\beta(\omega) = \beta(\omega) - \left( \beta_0 + \beta_1 \left[ \omega - \omega_0 \right] + \gamma P_P \frac{\omega}{\omega_0} - \frac{\omega_0}{2n_0 c} \frac{\rho}{\rho_{cr}} \frac{\omega_0}{\omega} \right),$$ (1)

where $\beta_0$ and $\beta_1$ are the propagation constant and inverse group velocity at the soliton centre frequency $\omega_0$, $\gamma$ is the pressure-dependent nonlinear fibre parameter[18] at $\omega_0$, $P_P$ the soliton peak power, $n_0$ the linear refractive index at $\omega_0$, $\rho$ is the plasma density and $\rho_{cr}$ the critical plasma density (at which the plasma becomes opaque at $\omega_0$) and $c$ is the speed of light in vacuum. The last two terms on the right-hand side of Eq. (1) originate from the Kerr effect and plasma formation[30]. While the Kerr effect is relevant at high frequencies due to its dispersive correction $\omega/\omega_0$, the plasma polarization, which goes as $\omega_0/\omega$, becomes dominant at low frequencies (inset in Fig. 4b). As a result, the phase matching point for DW emission in the UV can be tuned via the gas dispersion and, to some extent, the soliton peak power, while the plasma density has only a minor influence. In the MIR, however, the gas dispersion and the Kerr effect are comparatively weak, so that the phase matching is mainly governed by the plasma density created upon self-compression of the input pulse as it propagates along the fibre (Fig. 4a).

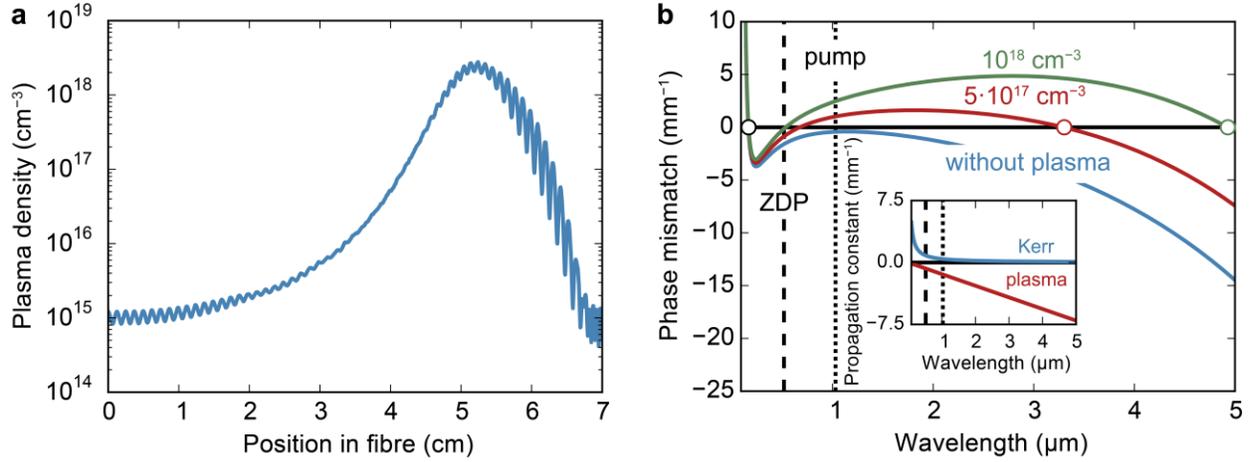

**Figure 4 | Plasma density evolution and DW phase matching. a,** Evolution of the plasma density with propagation distance for the simulation in Fig. 3. As the pulse propagates along the fibre, the carrier-envelope phase (global phase[31]) drifts and the field-dependent ionization rate dithers accordingly, creating ripples in the evolution of the plasma density. **b,** Phase mismatch $\Delta\beta$ according to Eq. (1) between a soliton with 1030 nm central wavelength and 1 GW peak power and DWs for the case when the fibre is filled with 4 bar of Ar and three different plasma densities (0, $5\cdot10^{17}$ and $10^{18}$ cm$^{-3}$). Phase matching to DWs is satisfied at zero phase mismatch (circles). The inset shows the contributions of the Kerr effect and plasma polarization to the nonlinear soliton propagation constant.

As illustrated in Fig. 4b, without any plasma contribution, phase matching to DWs is only possible in the DUV, on the opposite side of the ZDP[28]. For plasma densities greater than $1.2\cdot10^{17}$ cm$^{-3}$, two additional phase matching points appear, one of which lies in the MIR, progressively shifting to longer wavelength with increasing plasma density. For typical plasma densities reached during self-compression, e.g., $5\cdot10^{17}$ and $10^{18}$ cm$^{-3}$, phase matching to MIR DWs is satisfied at 3.3 and 4.9 μm respectively. While perfect phase matching is only achieved at zero phase mismatch, the shallow wavelength-dependence of the dispersion in the gas-filled kagomé-PCF guarantees cm-scale coherence lengths $L_{coh} = 2\pi/|\Delta\beta|$ for broadband DW emission in the UV and MIR (e.g., for the parameters given above, with a plasma density of $5\cdot10^{17}$ cm$^{-3}$, the coherence lengths for DW emission exceed 1 cm over bandwidths of 16 THz in the MIR (relative bandwidth of 18%) and 87 THz in the DUV (relative bandwidth of 5%)).

As phase matching to the MIR DW is not static, but varies dynamically with the plasma density (Fig. 4), the emission point of the MIR DW does not necessarily coincide with the point of maximum spectral broadening (Fig. 5). Instead, the interplay between spectral overlap with the pump soliton and phase matching governs the energy transfer to the MIR DW. In the limit of no pump depletion (a good approximation given the low conversion efficiency), the spectral power density of the generated DW takes the form[32]:

$$S_{\text{DW}}\left(\omega_{\text{DW}}\right) \propto S_{\text{P}}\left(\omega_{\text{DW}}\right)\text{sinc}^2\left(\frac{\Delta\beta\left(\omega_{\text{DW}}\right)L_{\text{gain}}}{2}\right),\qquad(2)$$

where $S_{\text{DW}}$ and $S_{\text{P}}$ are the spectral power densities of the DW and pump soliton and $\Delta\beta$ is the phase mismatch according to Eq. (1). For the MIR DW, the gain length $L_{\text{gain}}$ is limited to the mm-scale by temporal walk-off between the pump soliton and the generated DW (cf. Fig. 3c,d). The walk-off length based on group velocity mismatch for pulses with frequencies $\omega_1$ and $\omega_2$ and duration $\tau$ is given by $L_{\text{walk-off}} = \tau/|\beta_1(\omega_1) - \beta_1(\omega_2)|$, where $\beta_1 = \partial\beta/\partial\omega$ is the inverse group velocity.

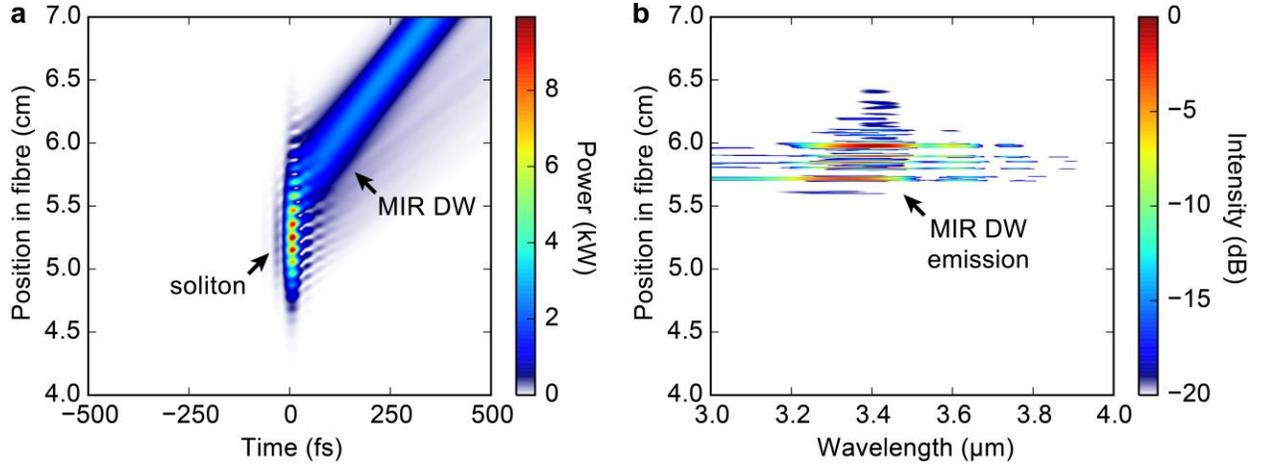

**Figure 5 | Dynamics of MIR DW emission. a,** Evolution of the temporal pulse shape with propagation distance, bandpass filtered between 3 and 4 µm, for the simulation shown in Fig. 3. **b,** Normalized spectral power density of a DW generated according to Eq. (2). The central wavelength of the soliton was determined from the first-order centroid of its spectrum. The gain length was defined as $L_{\text{gain}} = 1$ mm, based on the temporal walk-off between the soliton (central wavelength ~750 nm at the MIR DW emission point) and the MIR DW (~3.4 µm), which is 33 fs over a propagation distance of 1 mm.

Although the peak power of the soliton, in the spectral range between 3 and 4 µm, is strongest at the point of maximum spectral broadening, the MIR DW is emitted at a point slightly further along the fibre. At the point of maximum spectral broadening, the plasma density exceeds $10^{18}$ cm$^{-3}$, shifting the phase matching point of the MIR DW to beyond 5 µm, where the spectral overlap with the pump soliton is low. MIR DW emission mainly occurs for plasma densities of ~5·$10^{17}$ cm$^{-3}$, when the balance between phase matching and spectral overlap with the pump soliton is such that there is substantial parametric gain.

Experiments and numerical simulations show that DW emission is less efficient in the MIR than in the UV. There are two reasons for this. First, the optical shock effect strongly broadens the spectrum towards shorter wavelengths, whereas the spectral power of the soliton is weaker in the MIR as a result of less efficient spectral recoil[28]. Second, the high plasma density at the point of maximum compression impairs phase matching and prevents efficient transfer of energy to DWs in the MIR; this could be mitigated by balancing spectral broadening and plasma generation, e.g., by using two-colour pumping to maintain phase matching over a longer distance.

**Discussion**

Novel soliton-plasma interactions in the single-cycle regime permit emission of DWs in the MIR spectral region. The generated MIR DWs lie in the anomalous dispersion region, on the same side of the ZDP as the pump solitons, where phase matching is only possible through the transient plasma correction to the soliton propagation constant. Within a 4.7-octave-wide supercontinuum that reaches into the VUV, MIR DWs are generated in the wavelength range 3.2 to 4.7 μm. The results raise the prospect of a novel all-fibre-based source of ultrashort MIR pulses that complements state-of-the-art fibre laser technology[33,34], and could be useful for example for pumping soft-glass fibres for multi-octave MIR supercontinuum generation[35–37]. Even though the experimental parameters were not optimized for emission of UV DWs, up to 5.3 mW of average power was generated at ~200 nm, making the system ideal for spectroscopy applications or ultrafast pump-probe experiments from the VUV to the MIR[38–40].

**Methods—Experimental set-up**

The set-up consists of an ultrafast fibre laser, emitting 300 fs pulses (FWHM) at 1030 nm with 25 μJ energy at 151 kHz repetition rate, and two gas-filled ARR-PCF stages[41] (Supplementary Section 1). In the first stage the laser pulses are spectrally broadened by self-phase modulation (SPM) in a 19-cm-long single-ring ARR-PCF with 49 μm core diameter filled with 32 bar of Ar. Chirped mirrors provide a total of $-2100$ fs$^2$ group velocity dispersion (GVD) to compensate for the positive chirp induced by SPM and the material dispersion of the optical elements up to the input of the second fibre. Since a fixed negative GVD is introduced after the first fibre, the duration and phase of the compressed pulses can be fine-tuned by varying the input energy to the first fibre (and hence control the bandwidth generated through SPM), resulting in 27-fs-long (FWHM) pulses at the input to the second fibre. In the second stage, a 7-cm-long kagomé-PCF with 36 μm core diameter is used, filled either with 4 bar of Ar or 30 bar of Ne. The loss of this fibre is ~1 dB m$^{-1}$ at the pump wavelength (1030 nm), and below ~5 dB m$^{-1}$ over the wavelength range from 450 nm to 1.75 μm. Launch efficiencies into the fibres in both stages exceeded 90%.

## Methods—Experimental measurements

The output of the second fibre was collimated using an aluminum-coated off-axis parabolic mirror (OAPM) and either sent to a thermal power meter to measure the total output power (corrected for transmission through the magnesium fluoride window of the gas-cell and the reflectivity of the OAPM) or to two different spectrometers: a fibre-coupled silicon CCD spectrometer (intensity-calibrated between 200 and 1100 nm), or a monochromator equipped with a lead sulfide and a lead selenide detector (for the spectral range from 900 nm to 5 μm; not intensity-calibrated). The spectra of the two spectrometers were merged at a wavelength of 900 nm. During a wavelength scan with the monochromator, various order-sorting filters were used to suppress higher diffraction orders of the gratings. During supercontinuum generation, the residual pump signal is still strong. To verify that it does not cause artefacts in the measured spectrum, both fibres were evacuated to eliminate spectral broadening, and wavelength scans were performed over the entire spectral range (900 nm to 5 μm) using only one order-sorting filter at a time. In these scans, the strong pump signal was blocked below noise level (>60 dB rejection), and no signal was measured over the entire spectrum. During supercontinuum generation, further long-pass filters (cut-on wavelengths 3.1 μm and 3.6 μm) were inserted in addition to the order-sorting filters of the spectrometer to verify that the spectrum was free from artefacts.

For further measurements, the OAPM was replaced with a plano-convex magnesium fluoride lens. The DUV DW was spatially separated using a calcium fluoride prism, and its power measured with a silicon diode power meter and corrected for the transmission of the gas-cell window, lens, and prism. The MIR DW above 3.1 μm was filtered with two identical long-pass filters, and its power measured with a thermal power meter and corrected for the transmission of the gas-cell window, lens, and the two long-pass filters. For wavelengths longer than 3.1 μm, the optical near-field profile of the fibre mode at the output was imaged using an indium antimonide camera.

For measurements in the VUV, the output of the gas-cell was connected to a vacuum spectrometer equipped with an intensity-calibrated silicon photodiode. The spectral response of the system was determined from the transmission of the gas-cell window, the grating efficiency of the spectrometer, and the response of the photodiode.

## Methods—Numerical simulations

Pulse propagation was numerically simulated using a single-mode unidirectional field equation[30], including photoionization with the Perelomov, Popov, Terent'ev (PPT) ionization rates[42], modified with the Ammosov, Delone, Krainov (ADK) coefficients[43]. The dispersion of the fibre (36 μm core diameter, 200 nm core wall thickness, filled with 4 bar of Ar or 30 bar of Ne) was modelled using a simple capillary model[44], modified by a wavelength-dependent effective core radius that allows an accurate estimate of the modal refractive index at longer wavelength[45]. The *s*-parameter in this modified capillary model was set to $s = 0.08$, an optimum value that was determined by finite-element modelling of an idealized fibre with the same structural parameters. Since the parameters of the input pulse were measured by FROG, there were no free parameters in the simulations.

# Novel mid-infrared dispersive wave generation in gas-filled PCF by transient ionization-driven changes in dispersion

## SUPPLEMENTARY INFORMATION

F. Köttig[1*], D. Novoa[1], F. Tani[1], M. Cassataro[1], J. C. Travers[1,2], and P. St.J. Russell[1]

[1]Max Planck Institute for the Science of Light, Staudtstrasse 2, 91058 Erlangen, Germany

[2]School of Engineering and Physical Sciences, Heriot-Watt University, Edinburgh EH14 4AS, U.K.

*e-mail: felix.koettig@mpl.mpg.de

## S1. Experimental set-up

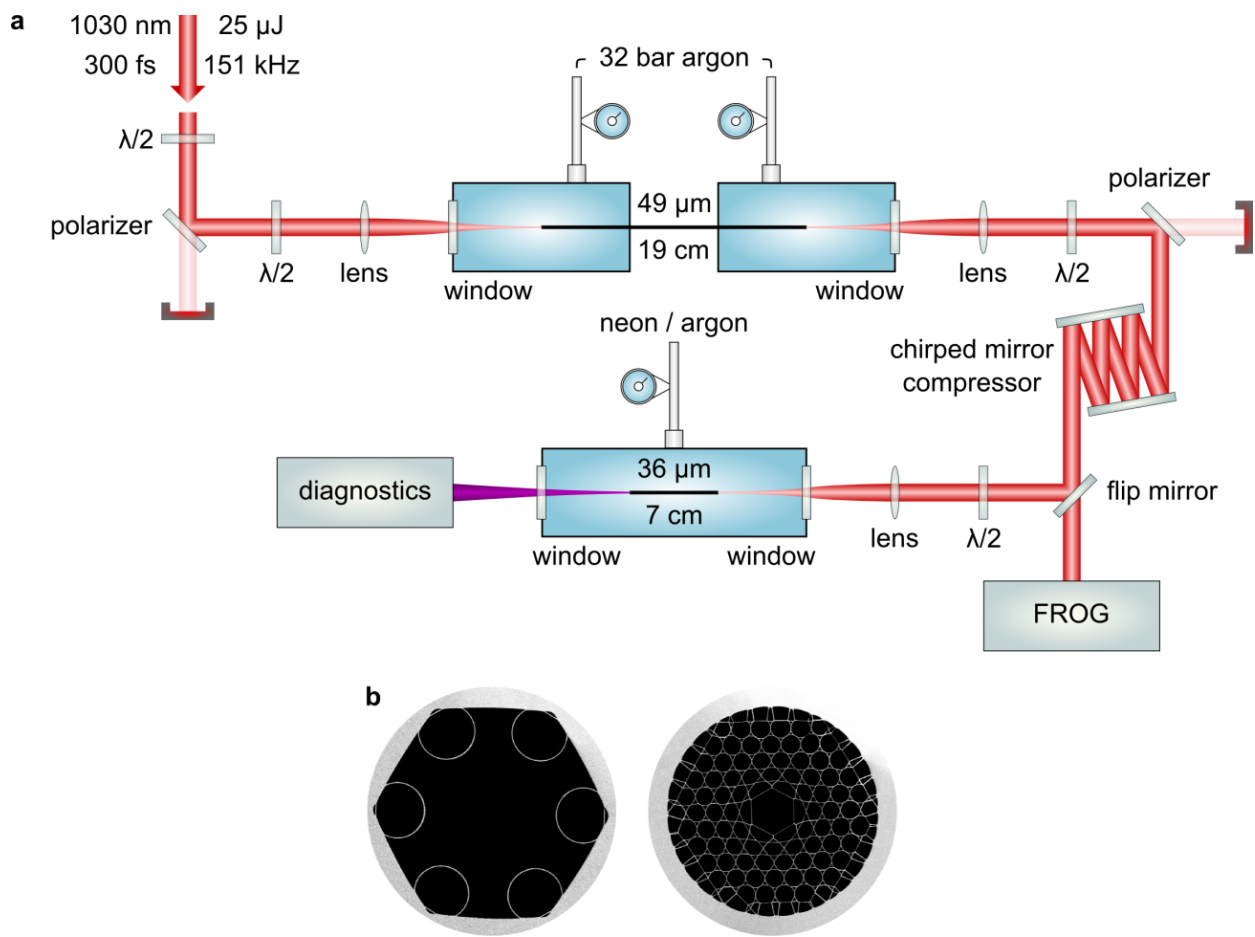

**Figure S1.1 | Experimental set-up. a,** The fibres are placed in gas-cells filled without pressure gradients. Half-wave plates and polarizers are used for power control. Half-wave plates before the fibres are used to

align the linear input polarization with the slightly birefringent axes of the fibres to optimize the polarization extinction ratio. **b,** Scanning electron micrograph of the single-ring (left) and kagomé-type (right) anti-resonant reflecting photonic crystal fibre (PCF) with 49 and 36 μm core diameter, used in the first and second stage, respectively. Both fibres have a loss of ~1 dB/m around the pump wavelength (1030 nm).

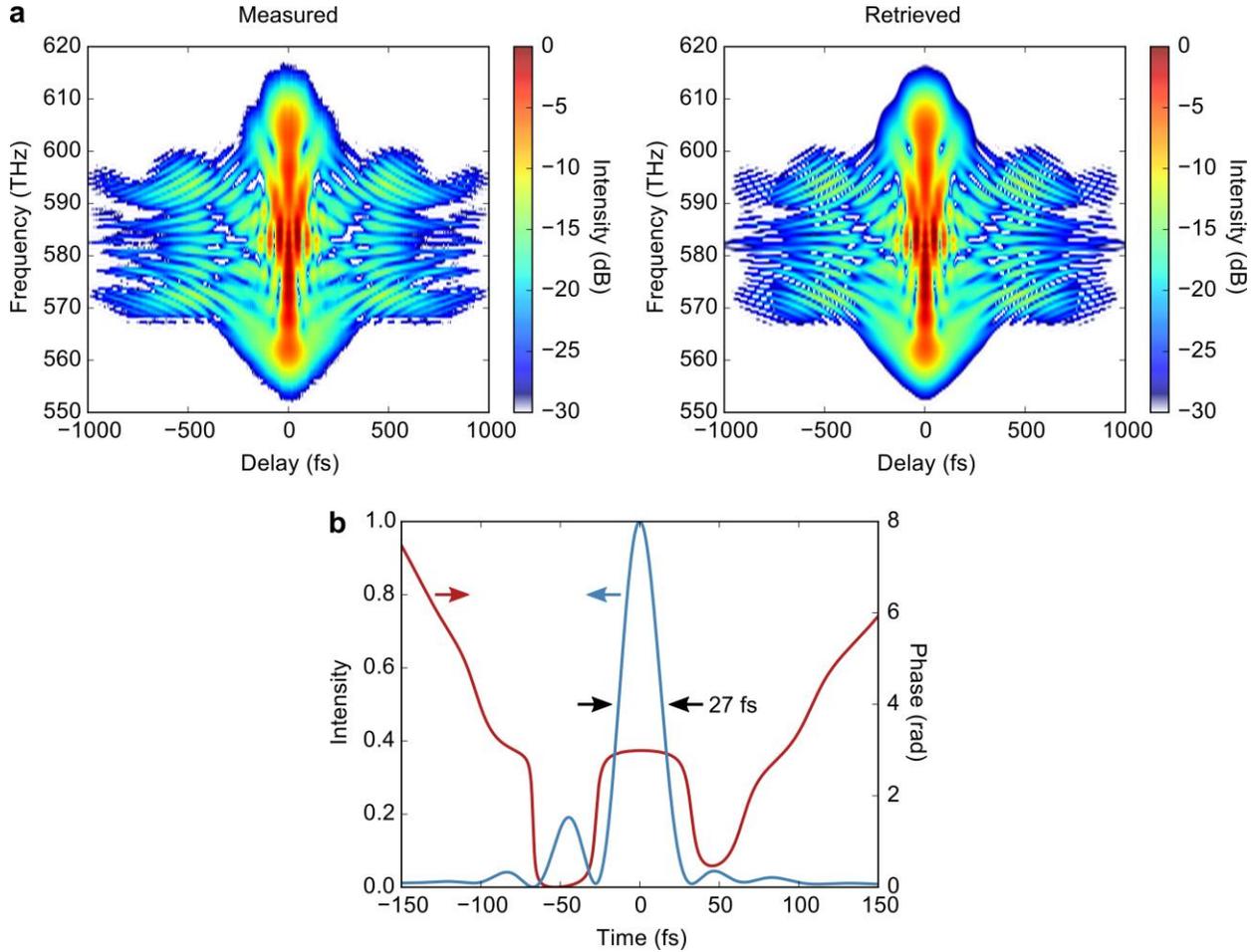

**Figure S1.2 | Input pulse to the second fibre. a,** Second harmonic generation frequency-resolved optical gating measurement of the pulse after compression. **b,** Pulse at the input of the second fibre (the dispersion of the half-wave plate, lens and gas-cell window before the input of the second fibre was added numerically). The full-width-half-maximum pulse duration is 27 fs.

## S2. Fibre dispersion

The fibres used in the experiment guide via anti-resonant reflection, and, despite their different structure, can be treated identically in terms of guided modes and dispersion. The modal refractive index $n$ of the (gas-filled) fibres can be approximated to good accuracy by that of a capillary fibre[1]:

$$n_{mn}\left(\lambda, p, T\right) = \sqrt{n_{\text{gas}}^2\left(\lambda, p, T\right) - \frac{\lambda^2 u_{mn}^2}{4\pi^2 a^2\left(\lambda\right)}}\,, \tag{S2.1}$$

where $\lambda$ is wavelength, $n_{\text{gas}}$ is the refractive index of the filling gas, $p$ is its pressure, $T$ its temperature, $a$ is the wavelength-dependent effective core radius of the fibre, and $u_{mn}$ is the $n$th zero of the $m$th-order Bessel function of the first kind. Through careful launch alignment, we effectively excite only the fundamental LP$_{01}$-like mode, for which $m = 0$ and $n = 1$. A wavelength-dependent effective core radius is required to accurately describe the modal refractive index at longer wavelength. It is given by[2]:

$$a\left(\lambda\right) = \frac{a_0}{1 + s\lambda^2/\left(a_0 h\right)}\,, \tag{S2.2}$$

where $a_0$ is the core radius of the fibre, $h$ is the core wall thickness, and $s$ is a model parameter. For the kagomé-PCF used in the experiment, $a_0 = 18\,\mu\text{m}$, $h = 200\,\text{nm}$, and the model parameter $s = 0.08$ was determined from finite-element modelling of an idealized fibre with the same structural parameters. Figure S2.1 shows the group velocity dispersion (GVD) of this fibre, calculated using Eq. S2.1 and S2.2.

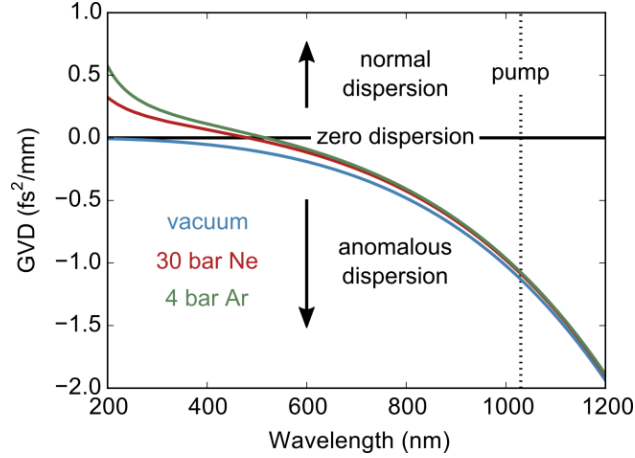

**Figure S2.1 | GVD of the kagomé-PCF with 36 µm core diameter.** The dispersion of the evacuated waveguide is anomalous everywhere, and the normal dispersion of the filling gas (30 bar Ne or 4 bar Ar) shifts the zero dispersion point into the visible.

### S3. Side-scatter measurement

Due to plasma generation inside the fibre, recombination luminescence is isotropically scattered through the side of the fibre, showing the characteristic emission lines of singly-ionized Ar (Fig. S3.1) or Ne. Distinct emission lines appear at the onset of soliton blue-shifting and mid-infrared (MIR) dispersive wave (DW) emission, indicating the strong correlation of these phenomena with the presence of a plasma inside the fibre.

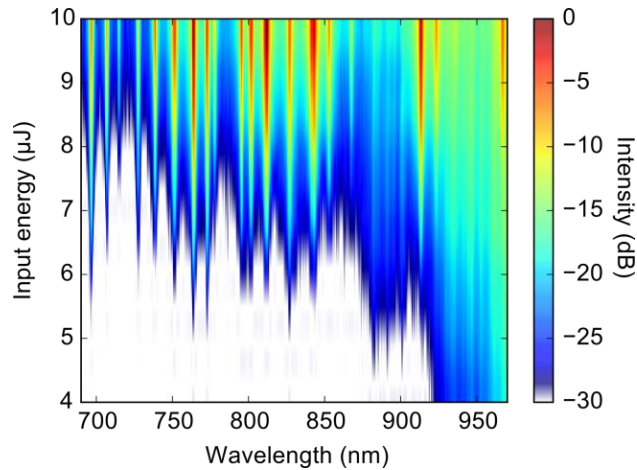

**Figure S3.1 | Side-scatter signal along the fibre.** Measured via 2f imaging onto a fibre-coupled silicon CCD spectrometer, when the fibre was filled with 4 bar of Ar. Part of the supercontinuum background comes from light that is reflected at the output window of the gas-cell and scattered inside. The side-scatter is measured at ~2 cm from the output of the fibre. With increasing input energy, the maximum temporal compression point (where the highest plasma densities are created) moves towards the input of the fibre, and the measured plasma fluorescence becomes stronger.

## S4. Ultrashort MIR pulses through DW emission

The emission of the MIR DWs is broadband, and numerical simulations predict a flat phase[3]. Considering the Fourier transform limit of the spectrum between 3 and 4 μm, few-cycle pulses in the MIR with peak power exceeding 20 kW generated through DW emission appear feasible (Fig. S4.1), and will be topic of future research.

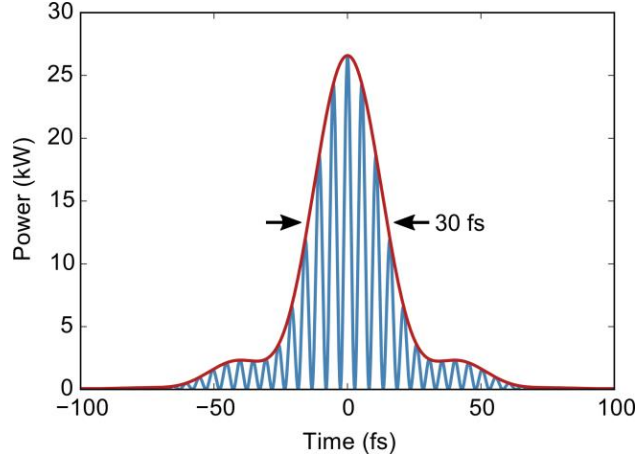

**Figure S4.1 | Transform limit of the MIR DW between 3 and 4 μm.** Fibre filled with 4 bar of Ar, 6 μJ input pulses. The transform limited pulse duration is 30 fs full-width-half-maximum.

## S5. Numerical simulation

The pulse propagation inside the fibre was numerically simulated using a single-mode unidirectional field equation[4]:

$$\frac{\partial \tilde{E}(z,\omega)}{\partial z} = i\left(\beta(\omega) - \beta_1 \omega\right)\tilde{E}(z,\omega) + i\frac{\omega^2}{2c^2 \varepsilon_0 \beta(\omega)}\tilde{P}_{NL}(z,\omega),$$
(S5.1)

where $\tilde{E}$ is the electric field in the frequency domain, $z$ is the position in the fibre (propagation distance), $\omega$ is frequency, $\beta$ is the propagation constant of the fibre mode (calculated using Eq. S2.1 and S2.2), $\beta_1$ is the inverse of the group velocity at the frequency of the pump, $c$ is the speed of light in vacuum, $\varepsilon_0$ is the vacuum permittivity, and $\tilde{P}_{NL}$ is the nonlinear polarization in the frequency domain given by

$$\tilde{P}_{NL}(z,\omega) = \mathcal{F}\left(\varepsilon_0 \chi^{(3)} E(z,t)^3 + P_{ion}(z,t)\right),$$
(S5.2)

where $\mathcal{F}$ denotes the Fourier transform, $\chi^{(3)}$ is the third-order nonlinear susceptibility of the gas, $t$ is time with respect to a reference frame propagating with group velocity $1/\beta_1$, and $P_{ion}$ is the plasma polarization. While the optical Kerr effect is governed by $\chi^{(3)} E(z,t)^3$, $P_{ion}$ is given by[5]

$$\frac{\partial P_{ion}(z,t)}{\partial t} = \frac{I_P}{E(z,t)} \frac{\partial \rho(z,t)}{\partial t} + \frac{e^2}{m_e} \int_{-\infty}^{t} \rho(z,t')E(z,t')\mathrm{d}t', \tag{S5.3}$$

where $I_P$ is the ionization potential of the gas, $\rho$ is the plasma density (calculated using the Perelomov, Popov, Terent'ev (PPT) ionization rates[6], modified with the Ammosov, Delone, Krainov (ADK) coefficients[7]), and $e$ and $m_e$ are the charge and mass of the electron. The numerical model (including ionization) was rigorously validated in previous work[8,9], and also yields excellent agreement between experimental results and simulations in this experiment.

Unlike in the experiment, ionization can be switched off in the numerical simulations (Fig. S5.1). In this case, no MIR DW is emitted since phase matching cannot be fulfilled. Also, the peak power of the self-compressed pulse is higher as there are no ionization-related losses (which are apparent in the measured transmission of the fibre, cf. Fig. 1c). The presence of plasma lowers the refractive index, leading to a blue-shift and acceleration of the pulse in the anomalous dispersion region (cf. Fig. S5.1a left, where the pulse curves to the left after the maximum temporal compression point). When ionization is switched off, however, the spectral recoil following the self-compression and DW emission effectively red-shifts the pulse, and it gets decelerated (cf. Fig. S5.1a right).

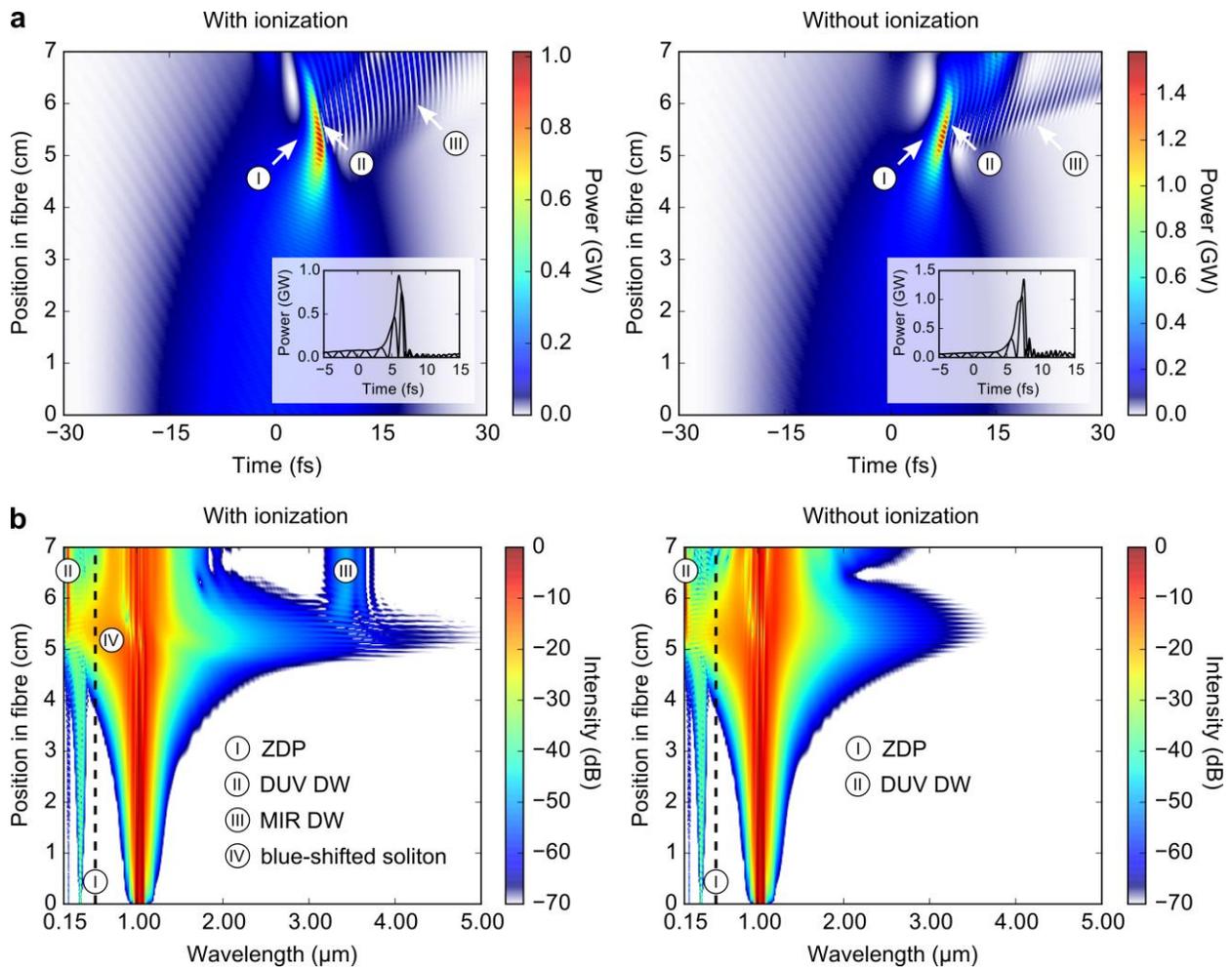

**Figure S5.1 | Simulated pulse propagation with and without ionization.** Fibre filled with 4 bar of Ar, 6 μJ input pulses (measured via frequency-resolved optical gating, Fig. S1.2b). **a,** Evolution of the temporal pulse shape with propagation distance. The inset shows the pulse at the maximum temporal compression point. (I)—maximum temporal compression point. (II)—shock front. (III)—deep ultraviolet DW. **b,** Evolution of the normalized spectral intensity (per unit wavelength) with propagation distance. The dispersion is anomalous on the long-wavelength side of the zero dispersion point ZDP (dashed line). The dynamics are very similar in both cases, yet no MIR DW is emitted when ionization is switched off.